\documentclass[a4paper,11pt]{article}
\usepackage{pos}
\usepackage{makecell}

\title{Overview of the Mini-EUSO $\mu s$ trigger logic performance}
 \ShortTitle{Overview of the Mini-EUSO $\mu s$ trigger logic performance}

\author*[a,b]{M. Battisti} \author[a,b]{M. Bertaina} \author[c,d]{M. Casolino} \author[e] {A. Belov}
\author[f]{F. Capel} \author[b]{M. Mignone} \author[a,b]{H. Miyamoto} \author[g]{L.W. Piotrowski, for the JEM-EUSO Collaboration}

\affiliation[a]{University of Turin}
\affiliation[b]{INFN Torino}
\affiliation[c]{INFN Roma Tor Vergata }
\affiliation[d]{RIKEN, Wako}
\affiliation[e]{MSU, Moscow State University}
\affiliation[f]{KTH Royal Institute of Technology}
\affiliation[g]{University of Warsaw}

\emailAdd{matteo.battisti@edu.unito.it}

\abstract{
Mini-EUSO is the first detector of the JEM-EUSO program deployed on the International Space Station (ISS). It is a wide field of view telescope currently operating from a nadir-facing UV-transparent window in the Russian Zvezda module on the ISS. It is based on an array of Multi-Anode Photomultipliers Tubes (MAPMTs) working in photon counting mode with a $2.5 ~ \mu s $ time resolution. Among the different scientific objectives it searches for light signals with time duration compatible to those expected from Extensive Air Showers generated by Extreme Energy Cosmic Rays (EECRs) interacting in the atmosphere. Although the energy threshold for cosmic ray showers is above $E>10^{21} eV$, due the constraints given by the size of the UV-transparent window, the dedicated trigger logic has been capable of the detection of other interesting classes of events, like elves, and ground flashers. An overview of the general performance of the trigger systems is provided, with a particular focus on the identification of classes of events responsible for the triggers.
}

\FullConference{37$^{\rm{th}}$ International Cosmic Ray Conference (ICRC 2021)\\
		July 12th -- 23rd, 2021\\
		Online -- Berlin, Germany}

\begin{document}
\maketitle

\section{Detector description}
\label{DETECTOR_DESCRIPTION}
\subsection{Instrument overview}
\label{Instrument_overview}
Mini-EUSO is a small telescope ($37 \times 37 \times 62~cm^3$) with an optical system made of two Fresnel lenses of 25 cm diameter each.
The focal plane is made of 36 Multi Anode Photomultiplier Tubes (MAPMTs Hamamatsu Photonics R11265-M64), for a total of 2304 pixels arranged in a square matrix. 
Four MAPMTs are grouped in a $2 \times 2$ matrix called Elementary Cell (EC). Each of the nine ECs shares a common high voltage power supply (HVPS) based on a Cockroft-Walton circuit. 
The system has an internal safety circuit that reduce the gain and the collection efficiency of the pixels in the EC unit in case of high current drain due to bright light.  This status is called \textit{cathode 2 mode}.
The focal plane is covered with a UV bandpass filters (2 mm of BG3 material), making the pixels mostly sensitive to $290~nm - 430~nm$  band, in the near UV. 
Mini-EUSO has a large Field of View of roughly $42 ^{\circ}$, with a pixel size on ground around $6.3~km$. The total footprint on ground is slightly less than $350 \times 350~km^2$ \cite{Launch_paper}.


\subsection{Data acquisition system}
\label{Data_acquisition_system}

Mini-EUSO features a multi-level data acquisition system able to store events in three different timescales at the same time, allowing the observation of the same event with different time resolutions. The slowest timescale ($40.96 ~ ms$) takes data continuously without a trigger system, and performs a monitor of the UV emission of the Earth. 
A frame of $40.96 ~ ms$ is called D3 Gate Time Unite, or simply D3 GTU.
The main focus of this contribution we will be on the $2.5~\mu s$ timescale and its trigger system, used for the detection of very fast events. A $2.5 ~ \mu s$ frame is called D1 GTU.
The L1 trigger code runs on a running buffer containing the $2.5 ~ \mu s$ data stream.
When an event is triggered the system stores 128 D1 GTUs, corresponding to $320 ~ \mu s$ of data, 64 GTUs before and 64 GTUs after the event. 
Mini-EUSO data acquisition system can store up to 4 consecutive events within the same slot of 5.24 seconds. 
After the fourth triggered event the system is in dead time and can not store any other data. The ability to save data is restored at the start of the next slot of $5.24s$. The advantages of this system are twofold. The first is the possibility to store consecutive events without dead time, provided that there are still available slots. 
The second advantage is that very long and very bright phenomena like, for example, lightning strikes, that could last up to 1 second or more, do not generate an enormous amount of triggers that would saturate the bandwidth.
Finally, it is possible to prevent given pixels from triggering through an external command sent by the on-board CPU. This is usually done only for the first orbit (from the switch-on moment until the first sunrise), when all the border pixels and two entire MAPMTs are masked.
The masking is needed because some border pixels are more prone to crosstalk and other non-linear effect, and in ultimate analysis, more prone to generate fake triggers.
An upgraded of the firmware, to store up to 8 triggers every 5.24 seconds, doubling the amount of data that it is possible to store with $2.5  ~\mu s$ resolution, and to implement the pixel mask for the entire acquisition session is currently under test.

\subsection{Trigger logic}
\label{Trigger_logic}

\begin{figure}[ht]
\centering
\includegraphics[width=.7\textwidth]{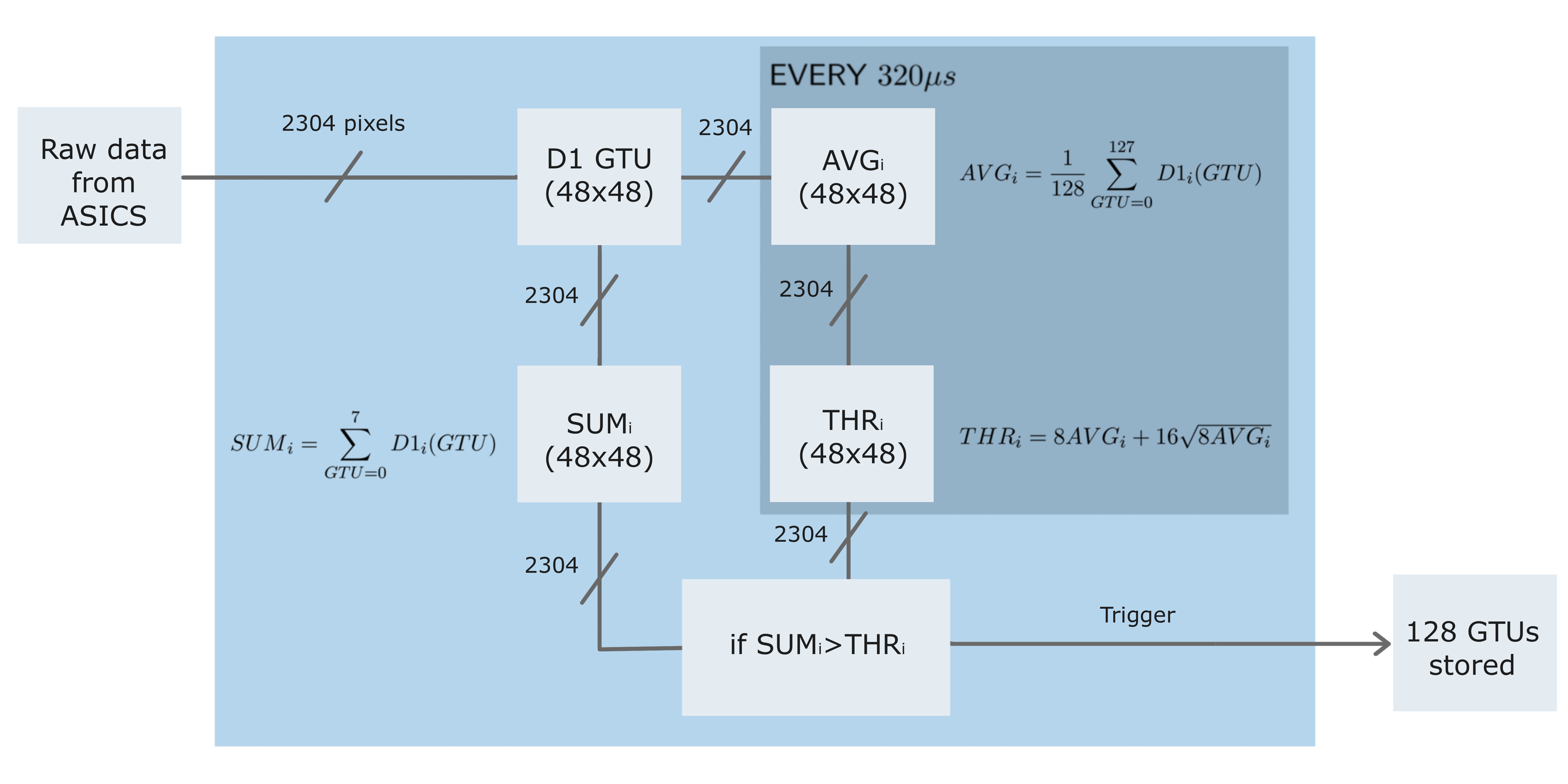}
\caption{Mini-EUSO trigger logic. Every $320 ~ \mu s$ the thresholds for each pixel are updated, according to the formulas in the darker box. In parallel, every GTU the integral over the last 8 GTUs is compared with the threshold for each pixel. If the sum is higher a trigger is issued}
\label{Mini-EUSO_L1_trigger_scheme}
\end{figure}

Given the large pixel size on ground and the fast sampling rate of Mini-EUSO, from the trigger point of view each pixel of Mini-EUSO has to be independent, with its own threshold; a single pixel above threshold is enough to issue a trigger. In fact a signal moving at the speed of light, like a EAS signal, takes slightly more than $20  ~ \mu s$ (8 GTUs) to cross 6.3 km, the field of view of a pixel. The logic therefore looks for an excess of signal over 8 consecutive GTUs in the same pixel, where the integration over 8 frames, namely $20  ~ \mu s$, enhances the signal to noise ratio.
The 2304 thresholds are updated every $320  ~ \mu s$, and set 16 $\sigma$ above the average value of the pixel over the previous slot of 128 D1 GTUs ($320  ~ \mu s$).
In Fig. \ref{Mini-EUSO_L1_trigger_scheme} is reported the scheme of the $2.5  ~ \mu s$ trigger logic.

\section{Analysis}
\subsection{Dataset}
\label{Dataset}
The following analysis is performed on a a dataset of 37.4 hours of data, coming from 31 different sessions; the excluded sessions were 
not performed in nominal working conditions. 


After each session a small portion of the data gathered are immediately downlinked to ground ( $\sim$ 40 minutes of data, that correspond to $\sim 20 \%$) while the remaining is stored in USB pen drives and delivered by the astronauts upon their return on ground. At the moment of writing more than $50 \%$ of data is still onboard the ISS. In the available dataset there are more than $5 \times 10^4$ L1 triggered events, an event being a collection of 128 frames with $2.5  ~ \mu s$ time resolution.

\subsection{D1 dead time}
\label{Dead_time}

\begin{figure*}[ht]
\centering
\includegraphics[width=.9\textwidth]{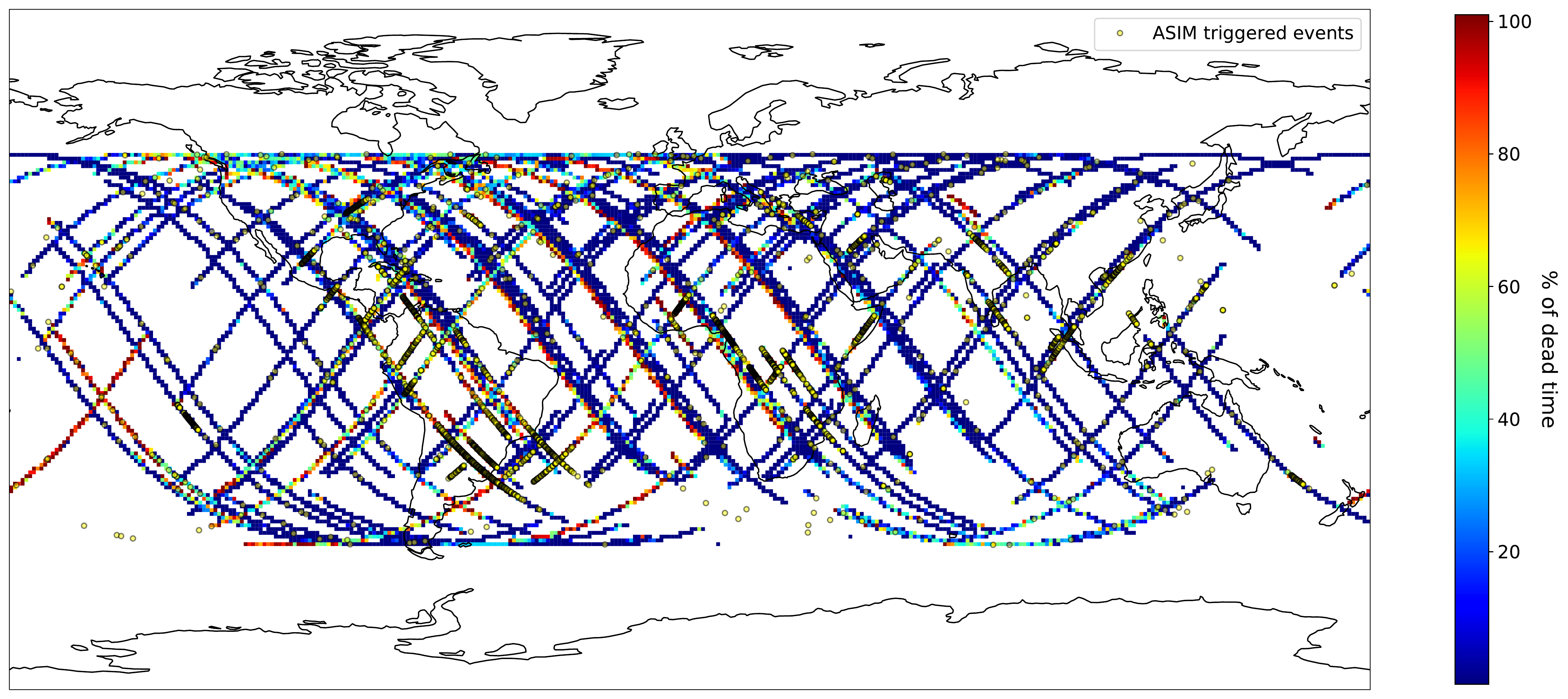}  
\caption{Fraction of dead time. The globe has been divided into a $1^{\circ} \times 1^{\circ}$ grid, the color represent the relative amount of dead time in each cell. The average dead time is $\sim  25\%$ . The black and yellow circles are triggers form ASIM detector}
\label{Relative_dead_time}
\end{figure*}

Figure \ref{Relative_dead_time} shows a map of the dead time of Mini-EUSO $2.5  ~ \mu s$ trigger logic. In black-yellow circles are reported the triggers from the Atmosphere-Space Interactions Monitor (ASIM), a detector installed outside the Columbus module of the International Space Station and devoted to the study of atmospheric events, namely thunderstorms and Transient-Luminous-Events (TLEs) \cite{ASIM}. The vast majority of its triggers are classified as lightning strikes.

 From the map it is clear that Mini-EUSO does not have higher dead time over cities and highly populated areas, since the adaptive threshold is fast enough to prevent static light sources from triggering. The higher concentration of dead time is usually due to thunderstorms that are usually triggered 4 times in just a few millisecond and quickly saturate the 4 available slots.

As previously mentioned, during the first orbit the border pixels and two MAPMTs are prevented from triggering to minimize the fake triggers due to non linear electric effect in these region. The trigger mask is not applied after the first orbit, resulting in an increase of dead time, from an average of 11\% in the first orbits to 29\% in all the others. The average dead time is $\sim  25\%$, 
which represent an acceptable level for Mini-EUSO, given the detector's characteristics. Future space-based and balloon-based detector (like EUSO-SPB2 or POEMMA) will work with a different logic for the data acquisition system, that will allow to obtain a lower dead time. In the meantime, the upgrade of the firmware presented in \ref{Data_acquisition_system} will significantly reduce this value for Mini-EUSO.

\subsection{Trigger performance}
\label{Trigger_results}
In order to evaluate the trigger performance, the  $2.5  ~ \mu s$ data have been analyzed through an offline algorithm that mimics the online trigger logic implemented in the FPGA. The values of the thresholds for each pixel have to be estimated from the data, since Mini-EUSO does not store them. In particular, the background for each pixel is assumed to be equal to the average over the first 32 GTUs ($80 ~ \mu s$) of each triggered packet. This is in general a good estimation since the 8 consecutive GTUs that issued the trigger are saved in GTUs 58-63 inside the 128 GTUs packet, therefore the first 32 GTUs contain only background. When long lasting events are triggered in more than one packet, the offline trigger code does not update the thresholds, using the ones computed for the previous packet.
The other option to estimate the background value would be to use the value of each pixel registered with the D3 timescale. The results of the two methods are quite similar in normal conditions, in the following analysis only the first method has been used, since is performs better when the background is changing quickly or in the presence of a transient event, but with the downside of being more susceptible to statistical fluctuations.  

Given the results of this offline algorithm, it is possible to produce a first categorization of the data, based upon the number of pixels over threshold in the entire packet of 128 GTUs. Notably, the trigger signal is issued when the first pixel is over threshold, all the others are irrelevant from the trigger point of view, but the presence of other pixels over threshold may give information on the event that produced the trigger.

Tables \ref{Trigger_table} and \ref{Pxl_OT_vs_packets} show that almost half of the triggered packets have only one pixel over threshold, as expected given the footprint on ground of a single pixel. For $\sim 18 \%$ of the triggered packets the algorithm was not able to recognise any pixel over threshold; this is due to the statistical fluctuations of the background over the first 32 GTUs of each packet, since an average made over 32 GTUs have a statistical fluctuation which is roughly double the one on 128 frames. To prove that, the same analysis has been repeated, setting the threshold to 12 $\sigma$ above the background and obtaining at least a pixel over threshold in $97 \%$ of the packets. This proves that in many of these packets there is at least one pixel with a signal just slightly below the threshold computed by the offline code, but above the real threshold used by the online trigger.
The remaining packets probably belong to some classes of events not always detected by the offline code, like the triggers issued by the sun rising, in which the light intensity grows constantly and quickly. These events might be difficult for the offline code that set the threshold based on the background $\sim 65 ~ \mu s $ before the triggered event, while the thresholds could be computed up to $320 ~ \mu s$ before, when the background is significantly lower.

\begin{table}[ht]
\begin{minipage}[b]{0.25\linewidth}
\centering
 \begin{tabular}{ |c|c| } 
 \hline\hline
  \makecell{Pixels \\ over \\ threshold}  & Frequency \\
 \hline\hline
 0 & 18.58 \%  \\ 
 1 & 45.73 \%   \\ 
 2 & 13.94 \%  \\ 
 3 & 5.25 \%  \\ 
 4 & 1.32 \%  \\ 
 5 & 0.62 \%  \\ 
 6 & 0.46 \%  \\ 
 \hline
\end{tabular}
    \caption{The numeric values of the first seven bins in Table \ref{Pxl_OT_vs_packets}}
    \label{Trigger_table}
\end{minipage}\hfill
\begin{minipage}[b]{0.7\linewidth}{
    \includegraphics[width=.99\textwidth]{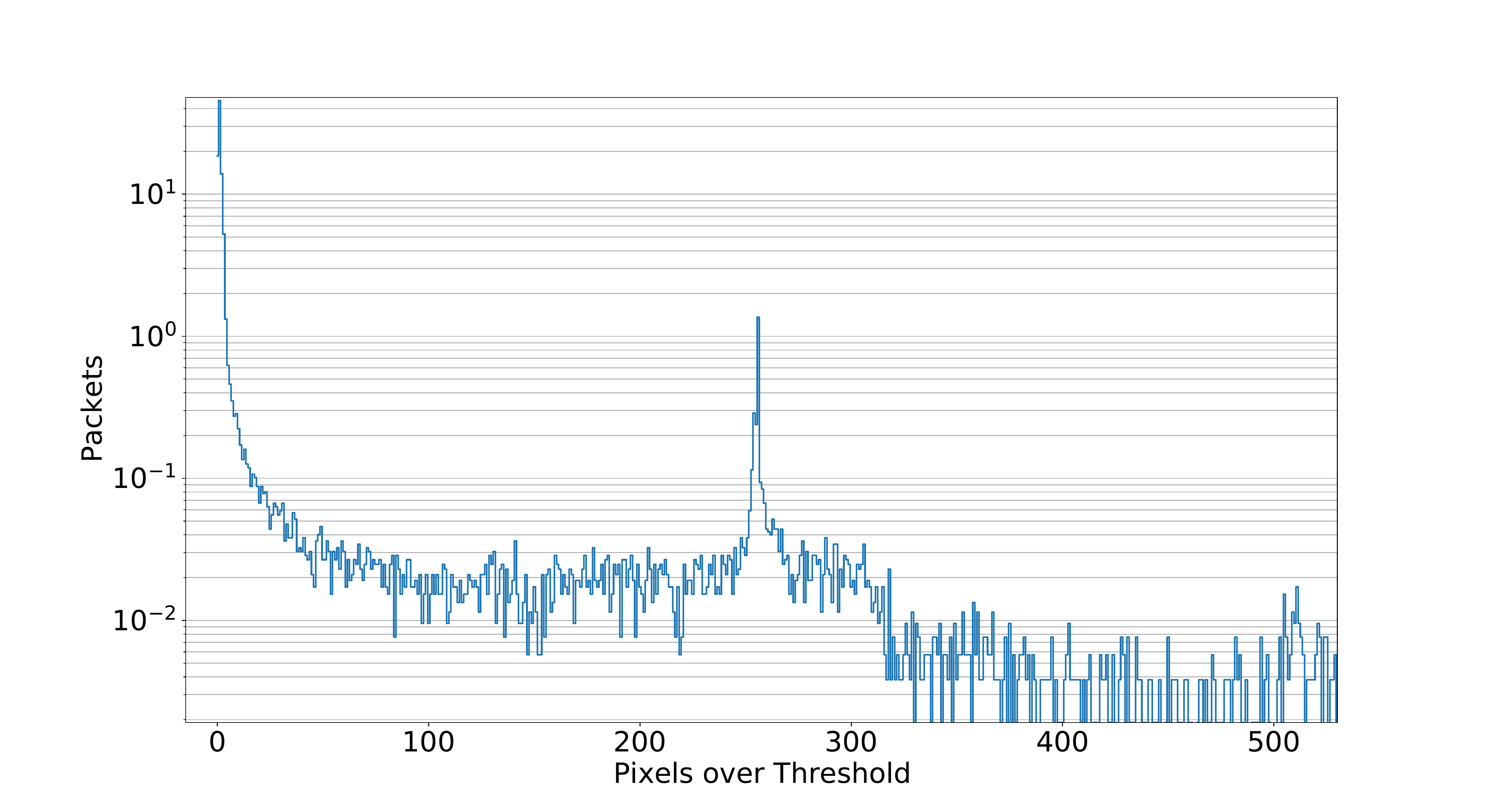}
    \caption{Frequency of triggered packets with a given number of pixels over threshold. The peak around 256 pixels corresponds to an entire EC switching back to nominal working conditions from \textit{cathode 2 mode}. A smaller peak is also present around pixel 512}
    \label{Pxl_OT_vs_packets}
    }
\end{minipage}
\end{table}

\begin{figure}[ht]
\centering
\includegraphics[width=1\textwidth]{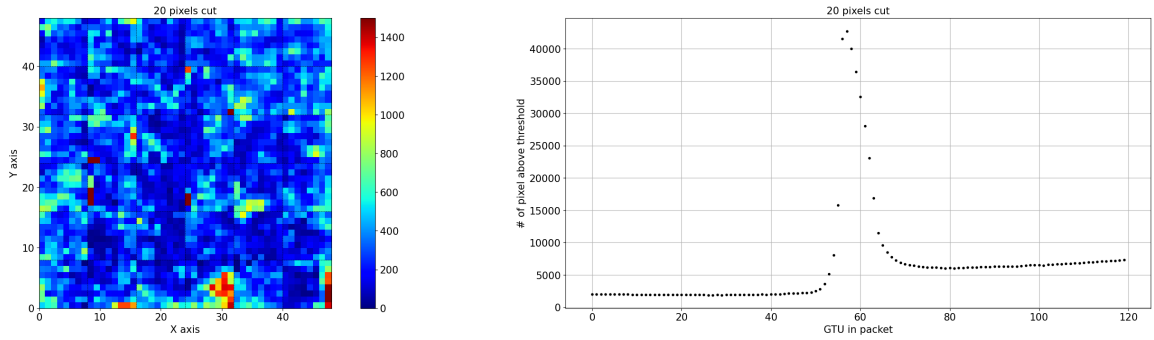}

\caption{\textbf{Left:} the number of GTUs over threshold for each pixel, considering only the packet with less than 20 pixels over threshold (46180 events, 88\% of the dataset). 12 border pixels (0.52\% of the pixels) account for 13.2\% of the total amount of pixels over threshold.
\textbf{Right:} the time position of the pixels over threshold inside the packet of 128 GTUs, each point is the sum over 8 consecutive GTUs: the huge peak centered at GTU 57 shows that the algorithm correctly recognises the events that issued the trigger, which are positioned at the center of the packet. A few long lasting events produce the tail, while almost no trigger are recognised before GTU 50}
\label{trigger_map-20pxl}
\end{figure}

The distribution of pixels over threshold is shown in Fig.\ref{trigger_map-20pxl} and presents a few hot spots, corresponding to border pixels more prone to generate fake triggers. On some occasions these pixels are responsible for a huge increase in the dead time, generating for example the orbit that in Fig.\ref{Relative_dead_time} starts from New Zealand and moves towards north-east, with an average dead time over 90\%. As already mentioned, a procedure to mask those pixels is currently under test, and will be implemented as soon as possible.

\section{Categories of triggered events}
\label{Catogories}
Being an orbital detector, Mini-EUSO detects a wide range of different events. A list of the most interesting or most common category of events is given in this section.
\begin{itemize}
\item \textbf{Atmospheric events:} lightning strikes are the brightest events detected by Mini-EUSO. Lightnings are best observed by Mini-EUSO with the longer timescale, but triggered also in the  $2.5 ~ \mu s$ time resolution. With this fast time resolution it is possible to probe the initial part of the developing of a lighting, and the signal appears over a large area of the focal plane where the brightness increases constantly, until the end of the 128 saved GTUs (Fig.\ref{Categories}, a).

\begin{itemize}
\item \textbf{Elves} are horizontally expanding, fast donut-shaped light emissions at the bottom ionosphere. The characteristic spatial dimension of elves are rings extending to a horizontal radius of $\sim 500 ~ km$. 
Mini-EUSO can provide high-speed UV imaging of elves, complementary to informations gathered by other instruments (i.e. ASIM, also on the ISS).
So far 17 elves have been detected (Fig.\ref{Categories}, b).
\end{itemize}
Other types of Transient Luminous Events (TLEs) are present in Mini-EUSO data and are currently under study.

\item \textbf{Ground flashers:} several triggers come from ground sources and present a light profile that lasts for tens on $\mu s$. These events are usually found near airports, and are probably produced by warning lights (Fig.\ref{Categories}, c). 


\item \textbf{EAS-like events:} The TUS detector has found several events with the shape and characteristics that resemble the ones expected from EASs, the most interesting being event TUS161003, over Minnesota. In Mini-EUSO data there are also several events with a lightcurve and time profile that matches the bi-gaussian shape expected from an EAS, with a faster rising and a slower decay, and the signal being confined in one or few neighbouring pixels. The vast majority of these events are detected near the location of airports and are triggered many times while moving in the focal plane, suggesting that airport's lights might be the sources of these events. In Fig. \ref{Minnesota_vs_Michigan} a comparison between TUS161003 and one of Mini-EUSO event is presented.

Through a comparison with a simulation, the cosmic origin has been excluded also for the only event with this characteristics found over the ocean and triggered only once (Fig.\ref{Pavel_event}). 

The research of these events is still ongoing.



\item \textbf{Direct cosmic rays} are low energy cosmic rays directly impinging on the detector. They are usually characterized by a signal that reach the maximum in only 1 GTU and then present an exponential decay. Direct cosmic rays are in general easy to recognise, and therefore do not represent a problem even considering that they represent a huge part of the $2.5 ~ \mu s$ dataset. They can appear with different shape on the focal plane, from a single bright pixel, to a blob or even a track.

\begin{figure*}[ht]
\centering
\includegraphics[width=0.9\textwidth]{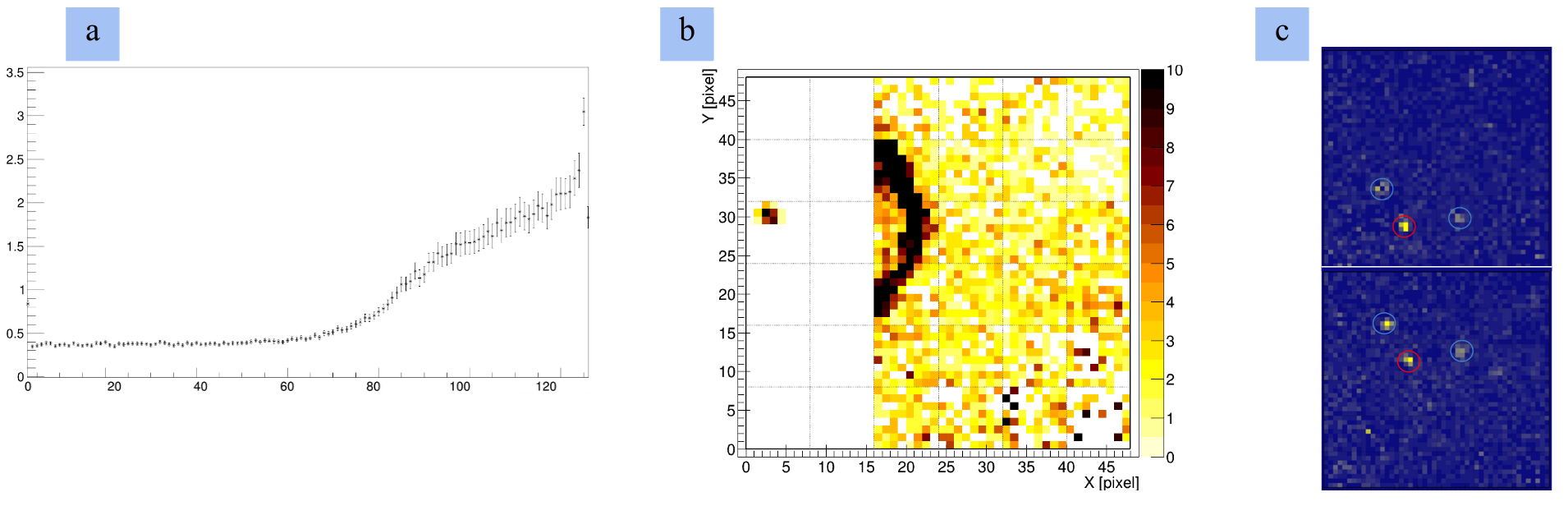}  
\caption{\textbf{a)}: Lightcurve of a ligthning event: after the trigger has been issued ($\sim$ GTU 64) the light  level increases constantly. \textbf{b)}: The bright disk of an elve. Mini-EUSO can observe the elve development as the ring expands in the field of view. \textbf{c)}: Two frames from two different events: the light sources inside the blue circles are static (cities) while the source inside the red circle is flashing and is what caused the trigger. 
}
\label{Categories}
\end{figure*}

\begin{figure*}[ht]
\centering
\includegraphics[width=1\textwidth]{Pictures/Pavel's_event.png}  
\caption{Focal plane view and lightcurve of Mini-EUSO signals and simulations. Left: Mini-EUSO event detected off the coast of Sri Lanka . Center and Right: EAS simulated through ESAF \cite{ESAF} with different energy and zenith angle. The simulation with $Z = 50^{\circ}$ and energy $5 \times 10^{21} ~ eV$  produce a footprint on the focal plane similar to the event but the lightcurve is too short, while the event at $Z = 80^{\circ}$ and energy $2 \times 10^{22} ~ eV$ correctly reproduce the lightcurve but has a different shape on the focal plane.}
\label{Pavel_event}
\end{figure*}

\begin{figure*}[ht]
\centering
\includegraphics[width=14cm,height=6cm]{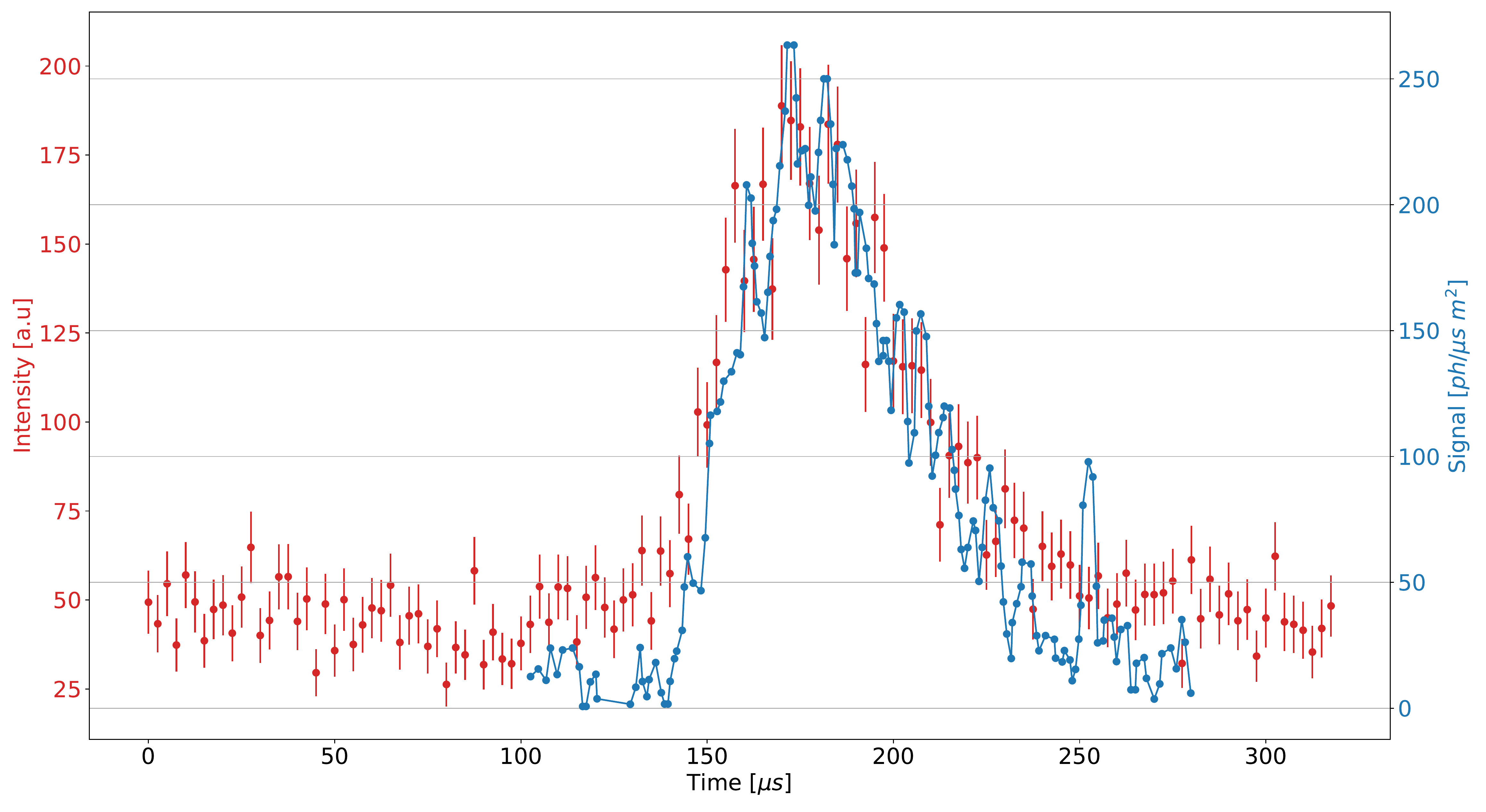}  
\caption{\textbf{Blue:} Lightcurve of TUS161003 event as seen by TUS. Even though the event has probably an anthropogenic origin, it presents all the features of an EAS signal. \textbf{Red:} An event detected by Mini-EUSO near lake Michigan. It presents the same time profile of the event seen by TUS even thoug it is $\sim$ 10 times brighter, assuming the detector efficiency quoted in \cite{Launch_paper}. It is triggered four times by Mini-EUSO, it is therefore produced by a ground source. The signal appears in an area near three small airports.}
\label{Minnesota_vs_Michigan}
\end{figure*}
\end{itemize}

\section{Conclusion}
Mini-EUSO $2.5 ~ \mu s$ trigger logic has proven to be working as expected. The fast adaptive thresholds prevents bright but static sources from triggering. The trigger logic and data acquisition system guarantee to keep the average dead time to slightly above 25\%. This is obtained without a second level trigger, and it is going to decrease once the firmware update  presented in \ref{Data_acquisition_system} will be installed. With the exception of a handful of pixels on the border, the distribution of pixels over threshold is quite uniform and it is possible to recognise different and interesting categories of events in the data.

%
%
%


\begin{thebibliography}{99}
\bibitem{Launch_paper} S. Bacholle \textit{et al.}; 
 2021 ApJS 253 36
\bibitem{ASIM} Neubert, T., Østgaard, N., Reglero, V. \textit{et al.}; 
Sci Rev 215, 26 (2019). 
\bibitem{TUS} AIP Conference Proceedings 566, 57 (2001); 
https://doi.org/10.1063/1.1378622
\bibitem{Minnesota_event} B.A. Khrenov \textit{et al.}; 
JCAP03(2020)033
\bibitem{ESAF} Berat \textit{et al.}, 
Issue 4, May 2010, Pages 221-247
\end{thebibliography}
\end{document}